\documentclass[10pt,aps,prb,twocolumn, nofootinbib, epsfig,showpacs,floatfix, bibliography]{revtex4-2}
\usepackage{amsmath}
\usepackage{xcolor}
\usepackage{url}
\usepackage{physics}
\usepackage{graphicx}
\usepackage{amsfonts}
\usepackage{mathrsfs}
\usepackage{capt-of}
\usepackage{amsmath}
\usepackage{amssymb}
\usepackage{float}
\usepackage{bbold}
 \usepackage[T1]{fontenc}
\usepackage[ttdefault=true]{AnonymousPro}
\usepackage[section]{placeins}
\usepackage[utf8x]{inputenc}
\usepackage{times}
\usepackage[colorlinks=true, linkcolor=blue, urlcolor=blue, citecolor=blue]{hyperref}
\makeatletter
\newcommand*{\rom}[1]{\expandafter\@slowromancap\romannumeral #1@}
\makeatother
\usepackage{multirow}
\hypersetup{
  colorlinks=true,
  citecolor=blue,
  linkcolor=blue,
  urlcolor=blue}

\begin{document} 

\title{ Work Statistics Under Quantum-Jump and Quench Dynamics in Monitored Ising Chains }
 
\author{Manali Malakar\textsuperscript{1,2}, Alessandro Silva\textsuperscript{1}}

\affiliation{\textsuperscript{1}International School for Advanced Studies (SISSA), Via Bonomea 265, 34136 Trieste, Italy\\
\textsuperscript{2}Institute for Cross-Disciplinary Physics and Complex Systems (IFISC) UIB-CSIC, Campus Universitat Illes Balears, 07122, Palma de Mallorca, Spain}

\date{\today}

\begin{abstract}
We investigate work statistics in monitored transverse-field Ising chains subjected to both a quantum quench of the transverse field and either stochastic quantum jumps or controlled measurement sequences. For generalized measurements,  we derive a trajectory-resolved generating function for work statistics in the two-point energy measurement scheme. Evaluating it using a fermionic Gaussian-state formalism, we show that, under stochastic jump dynamics, the work distribution crosses over from a comb-like structure to an essentially Gaussian form with shrinking sub-Gaussian tails, as the number of detection events grows. For controlled jump protocols, the energy added by each jump is constant when successive jumps are causally disconnected, but decreases and then saturates when they lie within each other’s light cone, leading to linear growth of average work in the former case and a transient sublinear regime followed by linear growth with a reduced slope in the latter. For monitored quenches, continuous observation washes out the fine structure of the isolated-quench distribution and again drives the statistics toward Gaussian behavior. Together, these results establish work statistics as a trajectory-resolved diagnostic of measurement-induced energy injection and of the emergence or breakdown of additivity in monitored many-body dynamics.
\end{abstract}

\maketitle
\section{Introduction}
\label{introduction}
Quantum thermodynamics is a rapidly developing field that seeks to extend the principles of stochastic
thermodynamics to far-from-equilibrium quantum systems~\cite{gemmer, anders, kosloff, talkner, silva}. In the quantum case, fluctuations of
thermodynamic quantities originate not only from the finite system size, making them more susceptible to thermal fluctuations, but also from quantum effects such as coherence, entanglement, and measurement backaction ~\cite{goold,mukamel,adesso,jacobs}. A primary task in this context is to define and measure work and heat in quantum processes, where energy exchanges are no longer continuous but instead occur via quantum jumps or coherent unitary drive. 
With appropriate definitions of work and heat~\cite{seifert},  fluctuation theorems such as the Jarzynski equality and the Crooks relation can still hold in the quantum domain, and one gains insight into energy exchange processes.\\~\\
The natural theoretical framework for describing the dynamics of quantum systems interrupted by measurements, whether in circuits or Hamiltonians, is the quantum trajectory formalism. In the case of weak measurements, the open-system dynamics at the level of individual realizations is governed by a stochastic Schrödinger equation (SSE)~\cite{milburn,piccito}. This approach provides a time-resolved picture of how work and heat are exchanged in each run of an experiment. Crucially, it enables defining thermodynamic quantities trajectory-wise and computing their full counting statistics by sampling many such trajectories~\cite{manzano, Elouard, malakar}. Recently this formulation has also been put to the test to discuss work statistics in postselected no-click trajectories, giving a meaning to this quantity in the non-Hermitian case~\cite{malakar}.\\~\\ 
Building on these foundations, studies of quantum-jump thermodynamics have explored the implications of detection events. The quantum-jump approach provides a natural way to identify energy flows: each detected jump carries away a quantum of energy, interpretable as heat lost or work done by the system depending on context~\cite{hekking}. Theoretical work has demonstrated that jump-resolved methods quantify dissipation in driven open systems, link jumps to entropy production, and yield moment-generating functions for energy exchange in general open dynamics~\cite{hekking,horowitz,mukamel,kemwing}. A recent experiment has reconstructed heat and work along individual quantum trajectories in superconducting qubits, validating the first law and underscoring the value of jump-resolved measurements for revealing rare events and modified fluctuation relations beyond ensemble averages~\cite{naghiloo}. However, most existing studies of monitored  dynamics in many-body systems have focused  on information-theoretic quantities such as entanglement (which leads to measurement-induced phase transitions~\cite{skinner}), rather than energetic fluctuations. An open question is therefore how multiple quantum jumps, separated in space and time, collectively shape work statistics in extended systems, with a focus on the conditions under which their correlations shape energy exchange.\\~\\
In this work, we address these questions by examining work statistics in a quantum Ising chain under full quantum-jump dynamics. Unlike prior "no-click" analyses that postselect no-jump trajectories and yield effective non-Hermitian evolution~\cite{malakar}, we incorporate stochastic jumps to probe real detection events and their impact on energy exchange. Using a fermionic Gaussian-state formalism, we determine the baseline work distribution for random quantum-jump trajectories as a function of measurement rate; with sufficiently many detections, the distribution develops an essentially Gaussian core with distinctly sub-Gaussian tails, and at large monitoring rates it approaches a fully Gaussian form. We then impose controlled jump sequences at specified times and locations to isolate the roles of jump timing and spatial separation, revealing strong, non-additive correlations within the light cone and independence outside. Finally, we combine continuous monitoring with a sudden unitary quench to uncover how coherent driving and measurement backaction interplay in shaping work statistics. This perspective makes the full work distribution, rather than only its mean, a diagnostic of how measurement backaction, quasiparticle propagation, and spatiotemporal jump correlations organize energy exchange in an extended quantum system.\\~\\
The rest of the paper is organized as follows. In Sec.~\ref{protocol}, we outline the quantum-jump protocol in the monitored transverse-field Ising model (TFIM) and derive the work generating function. Sec.~\ref{stochastic} presents results for stochastic jump dynamics, detailing work distributions and their sensitivity to monitoring strength and other system parameters. In Sec.~\ref{deterministic}, we explore controlled jump protocols, highlighting light-cone dependent correlations. Sec.~\ref{quench_dynamics} examines quench dynamics under monitoring, focusing on the interplay between coherent driving and stochastic monitoring. Finally, in Sec.~\ref{conclusion}, we summarize the findings and conclude.

\section{Measurement Protocol and Work Generating Function}
\label{protocol}
In this section, we establish the theoretical framework for analyzing work statistics in continuously monitored quantum systems, extending the conventional two-point measurement scheme to incorporate stochastic trajectories generated by quantum jumps.
\subsection{Stochastic Schrödinger Equation in a Monitored Quantum Ising Chain}
We begin by outlining the measurement protocol within the stochastic Schrödinger equation framework~\cite{milburn}. Let us now consider the quantum Ising chain
\begin{eqnarray}
\hat{H}_{\mathrm{QI}}=-J\sum\limits_{i=1}^{L}\hat{\sigma}^{x}_{i}\hat{\sigma}^{x}_{i+1}-h\sum\limits_{i=1}^{L}\hat{\sigma}^{z}_{i},
\label{QIC}
\end{eqnarray}
with periodic boundary conditions, subjected to local monitoring of its transverse spin component $\hat{\sigma}^{z}_{i}$, at measurement rate $\gamma$. In this work, we set \(J=1\), so energies are measured in units of \(J\). Such weak measurement processes are described by a positive operator-valued measure (POVM) with Kraus operators $\{\hat{A}_n\}$, obeying the completeness relation: $\sum_{n}\hat{A}^{\dagger}_{n}\hat{A}_n=\mathbb{1}$. For local monitoring, $n=\{r_{i}\}$, each operator is decomposed as: $\hat{A}_{n}=\bigotimes_{i=1}^{L} \hat{A}^{(r_i)}_{i}$, with $r_i \in \{0,1\}$ labeling the measurement results. With measurements performed over an infinitesimal time step $dt$, we evolve the system using the Kraus operators characterizing the quantum-jump protocol, given by
\begin{subequations}
\begin{align}
\hat{A}^{(0)}_{i} &= \hat{M}^{z}_{i-}+\sqrt{1-\gamma dt}\hat{M}^{z}_{i+},\\
\hat{A}^{(1)}_{i} &= \sqrt{\gamma dt}\hat{M}^{z}_{i+},
\label{Kraus}
\end{align}
\end{subequations}
where $\hat{M}^{z}_{i\pm}=(\mathbb{1}\pm \hat{\sigma}^{z}_{i})/2$ are projectors onto the eigenstates of $\hat{\sigma}^z_{i}$ with eigenvalues $\pm 1$. The full quantum-jump dynamics of the monitored system is then governed by the following stochastic Schrödinger equation~\cite{milburn,dalibard, daley,plenio}
\begin{eqnarray}
d|\psi_{t}\rangle&=&-i \hat{H}_{\mathrm {QI}}dt|\psi_{t}\rangle  - \frac{\gamma}{2}\sum \limits_{i}\bigg(\hat{M}^{z}_{i+}-\langle \hat{M}^{z}_{i+} \rangle_{t} \bigg) dt|\psi_{t}\rangle \notag\\
&+&\sum \limits_{i}\delta N^{i}_{t} \Bigg(\frac{\hat{M}^{z}_{i+}}{\sqrt{\langle \hat{M}^{z}_{i+} \rangle_{t}}}-1 \Bigg)|\psi_{t}\rangle,
\label{SSE}
\end{eqnarray}
where $\langle \hat{M}^{z}_{i+} \rangle_{t}=\langle \psi_{t}|\hat{M}^{z}_{i+}|\psi_{t}\rangle$, and $\delta N^t_{i} \in \{0,1\}$ denote independent Poisson counting processes with detection probability $p_{1}=\gamma dt \langle \hat{M}^{z}_{i+} \rangle_{t}$ (and $p_{0}=1-p_{1}$ for no detection).\\~\\
The individual trajectories generated by Eq.~(\ref{SSE}) allow the definition of thermodynamic observables, such as work, while explicitly accounting for the stochastic energy increments associated with jumps. However, in the no-click limit ($\delta N_i^t = 0$ for all $i$ and $t$), the SSE simplifies to a single evolution governed by a non-Hermitian Hamiltonian. This trajectory, though exponentially rare, provides potential analytical insights into measurement-induced effects. Indeed, the wave function evolution
\begin{eqnarray}\label{NC_wavefunction}
|\psi_t\rangle = \frac{e^{-i \hat{H}_{\mathrm{eff}} t} |\psi_0\rangle}{\| e^{-i \hat{H}_{\mathrm{eff}} t} |\psi_0\rangle \|},
\end{eqnarray}
with the effective non-Hermitian Hamiltonian
\begin{eqnarray}
\hat{H}_{\mathrm{eff}} = \hat{H}_{\rm QI} - i \frac{\gamma}{4} \sum_i \hat{\sigma}_{z}^i, 
\end{eqnarray}
can be studied analytically, where the imaginary term $-i (\gamma/4) \sum_i \hat{\sigma}_z^i$ arises from the continuous backaction.\\~\\

\subsection{Work Statistics Under Generalized Measurements}

In this paper, we define work using the standard two-point measurement protocol in isolated quantum systems, which consists of two projective energy measurements at the initial and final times, associated with the Hamiltonians $H_i$ and $H_f$, respectively. The work probability distribution is then given by~\cite{kurchan,Jarzinsky,Lutz}
\begin{eqnarray}~\label{ws}
P(W)=\sum_{n,m}p_i(n)p_f(m|n)\delta\bigg(W\!-\![E_f(m)\!-\!E_i(n)]\bigg),  
\end{eqnarray}
conditioned on the initial energy measurement, where $p_i(n)$ and $p_f(m|n)$ denote the probability of measuring the initial energy $E_i(n)$ and final energy $E_f(m)$. While the initial outcome is sampled from the thermal distribution as $p_i(n)=\exp[-\beta E_i(n)]/Z_i$ with partition function $Z_i$,  the conditional probability $p_f(m|n)$ depends on the details of the dynamics between the two measurements. For closed systems, evolving under a generic unitary operator $U_{t_f,t_i}$, the conditional probability is given by~\cite{Lutz}
\begin{eqnarray}
  p_f(m|n)=|\langle \psi_f(m)| U_{t_f,t_i}|\psi_i(n)\rangle|^2.  
\end{eqnarray}
To include generalized measurements during evolution, consider mid-process measurements with outcomes $\{r\}$ represented by operators $M_r$ satisfying $\sum_r M_r^\dagger M_r = \mathbb{1}$. For a single intermediate measurement with outcome $r$ at time $t_r$, the conditional probability reads~\cite{malakar}
\begin{eqnarray}
p_f(m|r,n)=\frac{|\langle \psi_f(m)|U_{t_f,t_r}M_rU_{t_r,t_i}|\psi_i(n)\rangle|^2 }{p(r|n)},   
\end{eqnarray}
where $p(r|n)=\langle \psi_i(n) |U^{\dagger}_{t_r,t_i}M^{\dagger}_r M_rU_{t_r,t_i}| \psi_i(n)\rangle$ is the probability of obtaining outcome $r$, given the initial energy $E_{i}(n)$. Here, we introduce the \textit{evolution operator} $T_{t_f,t_i}(r,t_r)=U_{t_f,t_r}M_rU_{t_r,t_i}$. Summing over all measurement records then yields the final transition probability
\begin{eqnarray}~\label{unconditionalwithmeas}
 p_f(m|n)&=&\sum_{\{r\}} p\big(r|n\big)\;p_f\big(m|r,n\big)=\nonumber \\
 && \sum_{\{r\}} |\langle \psi_f(m)|T_{t_f,t_i}\big(r,t_r\big)|\psi_i(n)\rangle|^2.
\end{eqnarray}
This expression can be further generalized to the case of $N$ intermediate measurements with outcomes $\{r_k\}$ at times $\{t_k\}$ ($k=1,\dots,N$), where the evolution operator takes the form
\begin{eqnarray}~\label{generalT}
 T_{t_f,t_i}(\{r_k,t_k\})=U_{t_f,t_N}M_{r_N}U_{t_N,t_{N-1}}
\dots M_{r_1}U_{t_1,t_i}.\notag
\end{eqnarray}
In this case,
\begin{eqnarray}
 p_f(m|n)=\sum_{\{r_k\}} |\langle \psi_f(m)|T_{t_f,t_i}\big(\{r_k,t_k\}\big)|\psi_i(n)\rangle|^2.
 \label{unconditionalwithmeas_v2}
\end{eqnarray}
Substituting Eq.~(\ref{unconditionalwithmeas_v2}) into Eq.~(\ref{ws}) yields the most general expression for the work statistics, i.e., the full work distribution in the presence of multiple generalized measurements.\\~\\
The work generating function
\begin{eqnarray}
 {\cal G}(u)=\int dW P(W) e^{-iWu},   
\end{eqnarray}
is finally given by the expression~\cite{malakar}
\begin{widetext}
\begin{eqnarray}\label{genfun}
{\cal G}(u)=\sum_{\{r_j\}} {\rm Tr}\left[ T^\dagger_{t_f,t_i}(\{r_j,t_j\})\;e^{-iH_f u}\; T_{t_f,t_i}(\{r_j,t_j\})\; e^{iH_iu}\frac{e^{-\beta H_i}}{Z_i}\right].
\end{eqnarray}
\end{widetext}
The normalization of the work distribution $P(W)$ (${\cal G}(u=0)=\int dW P(W)=1$) follows from the completeness relation 
\begin{eqnarray}\label{norma1}
 \sum_{\{r_j\}}  T^\dagger_{t_f,t_i}(\{r_j,t_j\})T_{t_f,t_i}(\{r_j,t_j\}) 
 ={\mathbb 1},
\end{eqnarray}
which is a direct consequence of the unitarity of $U$ and the completeness of $M_{r}$.
Setting $u=-i\beta$ instead, the Jarzynski equality reads
\begin{eqnarray}
{\cal G}(-i\beta)&=&\frac{{\rm Tr}\left[ \left(\sum_{\{r_j\}} T_{t_f,t_i}(\{r_j,t_j\})T^\dagger_{t_f,t_i}(\{r_j,t_j\}) \right) e^{-\beta H_f} \right]}{Z_i}\notag\\
&=&\frac{Z_f}{Z_i}=e^{-\beta \Delta F},\notag
\end{eqnarray} 
which is satisfied provided~\cite{Rastegin} 
\begin{eqnarray}\label{norma2}
 \sum_{\{r_j\}}  T_{t_f,t_i}(\{r_j,t_j\})T^{\dagger}_{t_f,t_i}(\{r_j,t_j\}) 
 ={\mathbb 1},
\end{eqnarray}
following from the requirement of unitality of the measurement operators, $ \sum_r M_rM^{\dagger}_r={\mathbb 1}$.\\~\\
For a normalized initial pure state $|\Psi_i\rangle$,  Eq.~(\ref{genfun}) can be written as

\begin{eqnarray}
\mathcal{G}(u) = \sum_{\{r_k\}}\langle \Psi_i|T^{\dagger}\big(\{r_k,t_k\}\big)e^{-iH_f u} T\big(\{r_k,t_k\}\big) e^{iH_i u}|\Psi_i\rangle.\notag
\end{eqnarray}
We may now rewrite the expression to be used within the formalism of the stochastic Schrödinger equation (SSE). Let us define the normalized SSE states as $|\Psi_{T,\textbf{r}}\rangle = T(\textbf{r})|\Psi_{i}\rangle/||T(\textbf{r})|\Psi_{i}\rangle||$, associated with the measurement record $\textbf{r}$. Taking the initial state $|\Psi_{i}\rangle$ as the ground state of $H_{i}$ with energy $E^{0}_{i}$, and averaging over trajectories, we obtain the generating function
\begin{eqnarray}
\mathcal{G}(u) = \sum_{\textbf{r}} p(\textbf{r})\langle \Psi_{T,\textbf{r}}|e^{-iH_f u}|\Psi_{T,\textbf{r}}\rangle e^{iE^{0}_i u},
\end{eqnarray}
where $p(\textbf{r})=||T({\textbf{r}})|\Psi_{i}\rangle||^2$ is the probability of observing the record $\textbf{r}$.

\section{Work Statistics under Stochastic Jump Dynamics}
\label{stochastic}
\begin{figure*}
\centering
\includegraphics[width=0.9\textwidth]{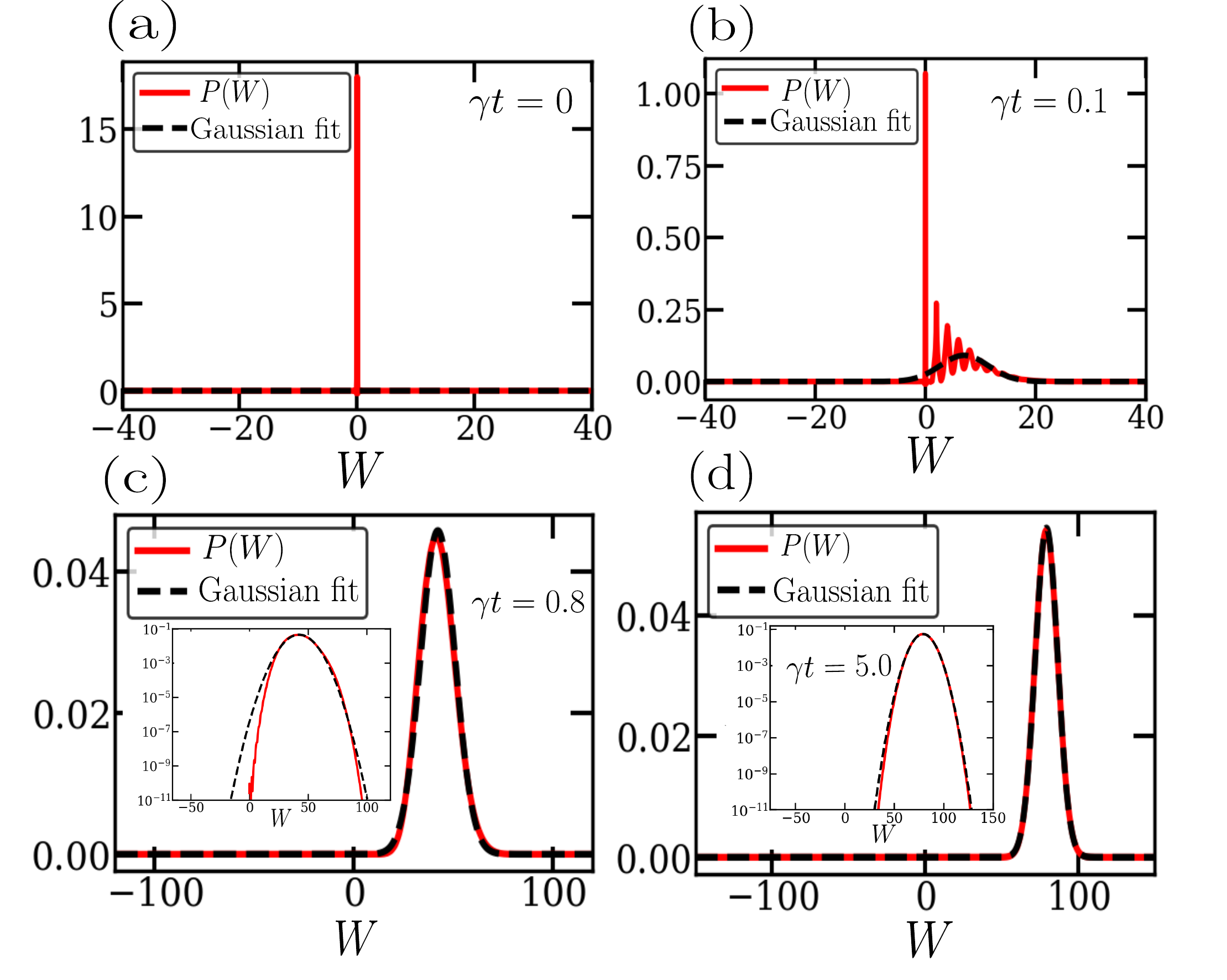}
\caption{Work distribution $P(W)$ under stochastic quantum-jump dynamics in a monitored
transverse-field Ising chain without quench ($H_i = H_f$), for system size $L=80$, transverse field $h = 0.4$, averaged over $200$ stochastic realizations. Solid red lines show the numerically constructed $P(W)$ and black dashed curves are Gaussian fits with the same mean and variance. (a) No monitoring ($\gamma t = 0$): the characteristic function reduces to $G(u)=1$ and the work distribution collapses to a single delta peak $P(W)=\delta(W)$ at $W=0$. (b) Weak monitoring ($\gamma t = 0.1$): rare jumps create a comb of narrow side peaks at $W>0$ on top of the dominant $W=0$ peak, reflecting discrete excitation pathways and residual coherences. (c) Intermediate monitoring ($\gamma t = 0.8$): many bounded mode-resolved energy kicks accumulate, producing an apparently Gaussian core on a linear scale; the inset shows on a logarithmic scale that the tails are sub-Gaussian, decaying faster than the best Gaussian fit. (d) Strong monitoring ($\gamma t = 5.0$): frequent jumps strongly dephase the dynamics so that higher cumulants are suppressed and $P(W)$ becomes effectively Gaussian, both in the bulk and in the logarithmic-scale inset, consistent with a central-limit mechanism for the summed work increments.}
\label{Fig1}
\end{figure*}
The purpose of this section is to numerically examine the effects of stochastic quantum jumps on the work statistics of a monitored Ising chain, exploring how the work probability distribution varies with the number of jumps. Under the stochastic jump evolution described in Eq.~(\ref{SSE}), the quantum state remains pure, and since the model is quadratic in free fermions, we can exploit Gaussianity to represent the full quantum evolution using the two-point correlation matrix. Under stochastic jump evolution, the
conditional state \(|\Psi_{T,r}\rangle\) remains pure and Gaussian.
Therefore, each trajectory is completely specified by the generalized
Nambu correlation matrix
\begin{equation}
\mathcal C^{(r)}
=
\begin{pmatrix}
C^{(r)} & F^{(r)} \\
F^{(r)\dagger} & \mathbb 1-C^{(r){\rm T}}
\end{pmatrix},
\end{equation}
where
\[
C_{ij}^{(r)}
=
\left\langle
\hat{c}_i^\dagger \hat{c}_j
\right\rangle_r,
\;\; \text{and}\;\;
F_{ij}^{(r)}
=
\left\langle
\hat{c}_i^\dagger \hat{c}_j^\dagger
\right\rangle_r.\]
The trajectory-resolved characteristic function is
\begin{equation}
\mathcal{G}^{(r)}(u)
=
e^{iE_i^0u}
\langle \Psi_{T,r}|e^{-iu\widehat H_f}|\Psi_{T,r}\rangle .
\end{equation}
We write the final quadratic Hamiltonian in terms of the Nambu spinor $\Gamma=\left(\hat{c}_1,\ldots,\hat{c}_L,\hat{c}_1^\dagger,\ldots,\hat{c}_L^\dagger\right)^{T}$, obtaining the following BdG form
\begin{equation}
\widehat H_f
=\Gamma^\dagger \mathcal H_f \Gamma
\end{equation}
up to an additive constant, which is absorbed into the reference energy. Diagonalizing the BdG matrix,
\begin{equation}
U_f^\dagger \mathcal H_f U_f
=
\mathcal E_f ,
\qquad
\mathcal E_f
=
{\rm diag}(\varepsilon_1,\ldots,\varepsilon_{2L}),
\end{equation}
and defining the final BdG spinor
\begin{equation}
\widetilde{\Gamma}
=
U_f^\dagger\Gamma ,
\end{equation}
we obtain
\begin{equation}
\widehat H_f
=\widetilde{\Gamma}^{\dagger}
\mathcal E_f
\widetilde{\Gamma},
\end{equation}
The correlation matrix in the same final BdG basis is
\begin{equation}
\widetilde{\mathcal C}^{(r)}
=
U_f^\dagger \mathcal C^{(r)} U_f .
\end{equation}
Thus
\begin{equation}
\mathcal{G}^{(r)}(u)=e^{iE_i^0u}
\left\langle \Psi_{T,r}\left|
e^{-iu\widetilde{\Gamma}^{\dagger}
\mathcal E_f
\widetilde{\Gamma}}
\right|\Psi_{T,r}\right\rangle .
\end{equation}
Since \(|\Psi_{T,r}\rangle\) is a pure Gaussian state,
\(\widetilde{\mathcal C}^{(r)}\) is a rank-\(L\) projector. We write
\begin{equation}
\widetilde{\mathcal C}^{(r)}
=
Q_rQ_r^\dagger ,
\qquad
Q_r^\dagger Q_r=\mathbb 1_L ,
\end{equation}
where the columns of \(Q_r\) span the occupied Gaussian subspace in the
final BdG basis. The single-particle action of the operator
\(e^{-iu\widetilde{\Gamma}^{\dagger}\mathcal E_f\widetilde{\Gamma}}\)
is
\begin{equation}
D_f(u)
=
e^{-iu\mathcal E_f}.
\end{equation}
Therefore, the Gaussian matrix element is the determinant of the
single-particle evolution projected onto the occupied subspace,
\begin{equation}
\left\langle \Psi_{T,r}\left|
e^{-iu\widetilde{\Gamma}^{\dagger}
\mathcal E_f
\widetilde{\Gamma}}
\right|\Psi_{T,r}\right\rangle
=
\det\!\left[
Q_r^\dagger D_f(u)Q_r
\right].
\end{equation}
Using \(\widetilde{\mathcal C}^{(r)}=Q_rQ_r^\dagger\) and
\(\det(\mathbb 1+AB)=\det(\mathbb 1+BA)\), this can be rewritten as
\begin{align}
\det\!\left[
Q_r^\dagger D_f(u)Q_r
\right]
&=
\det\!\left[
\mathbb 1_L
+
Q_r^\dagger
\left(D_f(u)-\mathbb 1_{2L}\right)
Q_r
\right]  \nonumber \\
&=
\det\!\left[
\mathbb 1_{2L}
+
\left(D_f(u)-\mathbb 1_{2L}\right)
Q_rQ_r^\dagger
\right]  \nonumber \\
&=
\det\!\left[
\mathbb 1_{2L}
-
\widetilde{\mathcal C}^{(r)}
+
D_f(u)\widetilde{\mathcal C}^{(r)}
\right].
\end{align}
Substituting \(D_f(u)=e^{-iu\mathcal E_f}\), we finally obtain
\begin{equation}
\mathcal{G}^{(r)}(u)=e^{iE_i^0u}\det\!\left[\mathbb 1_{2L}
-\widetilde{\mathcal C}^{(r)}+e^{-iu\mathcal E_f}
\widetilde{\mathcal C}^{(r)}
\right].
\label{eq:single_traj_G_fullcorr}
\end{equation}
This is the full-correlation Gaussian expression used in the numerical
calculation. It keeps the off-diagonal correlations generated by the
stochastic jump dynamics. 
\\~\\
Subsequently, to compute the trajectory-averaged ${\cal G}(u)$, we repeat the above for $l=1, \dots, R$ realizations, generate the trajectories $\{r_{l}\}$ and compute the following.
\begin{eqnarray}\label{g_avg}
{\cal G}(u) = \frac{1}{R}\sum_{l=1}^{R}  {\cal G}^{(r_l)}(u)  
\end{eqnarray}
To analyze the work distribution $P(W)$ numerically, we evaluate the inverse Fourier transform
\begin{eqnarray}\label{ifft}
P(W) = \frac{1}{2\pi}\int^{u_{\rm max}}_{-u_{\rm max}} du\; e^{i u W}\; {\cal G}(u).
\end{eqnarray}
We discretize \(u\) on a symmetric, uniformly spaced grid \(u_j\in[-u_{\max},u_{\max})\) with \(N_u=2^{15}\) points. The grid spacing is therefore \(\Delta u=2u_{\max}/N_u\). To mitigate ringing effects arising from the finite \(u\)-window, we multiply the characteristic function with a Gaussian factor, \(\exp\left(-u^2/2\sigma_u^2\right)\),
before performing the inverse Fourier transform. In the numerical calculations, we use broad \(\sigma_u\simeq 100\),  and choose \(u_{\max}\) in the range \(400\text{--}600\). This ensures that the integration window remains much wider than the Gaussian taper, \(u_{\max}\gtrsim 4\sigma_u\), so that edge effects are suppressed while the reconstructed work distribution is minimally distorted.
\\~\\
We analyze the dependence of the work distribution $P(W)$ on the cumulative measurement strength $\gamma t$ during an evolution of length $t$, and hence on the expected number of jumps. For simplicity and to focus solely on measurement effects, we choose $H_i=H_f$. As shown in Fig.~\ref{Fig1}(a), in the baseline case $\gamma=0$, the characteristic function reduces to ${\cal G}(u)=1$, and the work distribution collapses to a single delta peak $P(W)=\delta(W)$~\cite{Lutz}. For small but nonzero $\gamma t$, quantum jumps occur only occasionally along each trajectory. Each jump stochastically redistributes the occupations of the single-particle modes, creating excitations with specific energies $\{\mathcal{E}\}$. With few such events, only a limited set of excitations contributes, and the coherences between them are largely preserved. The resulting work probability density, shown in Fig.~\ref{Fig1}(b), appears as a series of narrow spikes: a dominant central peak at $W=0$ for trajectories with no jump plus small side peaks at $W>0$ located near integer multiples of the mode energies. This pattern reflects the discrete energy spectrum and residual interference effects.\\~\\
At intermediate $\gamma t$, jumps are more frequent, and measurement-induced dephasing becomes large enough to suppress coherent features. The work then becomes the sum of many small, mode-resolved increments bounded by the single-particle bandwidth. The aggregate produces an apparently Gaussian core in the linear plot of Fig.~\ref{Fig1}(c), as expected from the central-limit theorem (CLT). However, the inset on a logarithmic scale reveals that the tails decay faster than the Gaussian reference with the same mean and variance, which we refer to as sub-Gaussian behavior for sums of bounded contributions~\cite{hoeffding}. The left–right asymmetry is also expected in the present setup. Since the system is initially prepared in the ground state and $H_i=H_f$, work cannot be negative: in the absence of jumps, one has $W=0$, while each jump can only inject energy into the system. As a result, the distribution is bounded on the left by $W=0$ and extends only toward positive $W$, which makes it broader on the right than on the left. In Fig.~\ref{Fig1}(d), at sufficiently large $\gamma t$, frequent jumps strongly dephase the dynamics; the total work accumulates as the sum of many small energy increments. In the high-rate regime, higher-order cumulants are suppressed and the variance dominates. The characteristic function is therefore well captured by a quadratic cumulant expansion, making $P(W)$ effectively Gaussian, not only in the bulk but also on a logarithmic scale over the plotted range (see inset). This behavior aligns with the standard diffusion approximation for rapid jump processes (like a drift-diffusion limit described by the Fokker-Planck equation), although our conclusions do not rely on invoking that continuum limit~\cite{kurtz}.\\~\\
Notably, we still observe a Gaussian form of $P(W)$, although the jump process is an unraveling of a Markovian (Lindblad) master equation, successive work increments are not strictly independent; rather, they have short-range correlations due to measurement-induced dephasing and mixing. The reason is that, for such additive functionals of Markov processes, a CLT holds even though successive increments are not strictly independent: after centering and appropriate scaling, the sum is asymptotically normal and the cumulant generating function is quadratic near $u=0$, which is why the core of $P(W)$ remains Gaussian even though jumps are correlated. Therefore, our findings are consistent with established CLTs for additive functionals of Markov processes, and with the quantum-trajectory picture underlying Markovian monitored dynamics~\cite{plenio,kipnis,jones}.\\~\\
To further quantify how the non-Gaussian part of the work distribution changes with measurement strength, we compare the numerical distribution \(P(W)\) with its Gaussian reference \(P_G(W)\), constructed from the same mean and variance as \(P(W)\). We focus on the probability weight outside the central region of the distribution and define
\begin{equation}
D_{\rm tail}^{(2\sigma)}=
\int_{|W-\langle W\rangle|>2\sigma_W}dW\,
\left|P(W)-P_G(W)\right|.
\label{eq:Dtail_2sigma}
\end{equation}
This quantity measures the absolute difference between $P(W)$ and its Gaussian reference outside the interval
\([\langle W\rangle-2\sigma_W,\langle W\rangle+2\sigma_W]\). It therefore probes the rare-event part of the work distribution, where deviations from Gaussian behavior are most visible.
\begin{figure}[h!]
\centering
\includegraphics[width=0.85\linewidth]{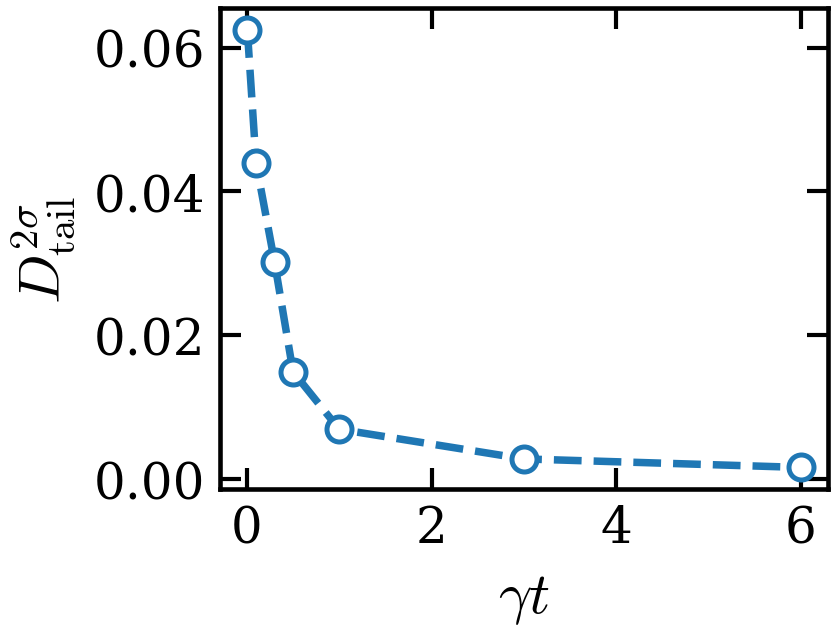}
\caption{Variation of the tail deviation \(D_{\mathrm{tail}}^{(2\sigma)}\) with the cumulative monitoring strength \(\gamma t\) at transverse field \(h=0.6\). The decrease of \(D_{\mathrm{tail}}^{(2\sigma)}\) indicates that the work distribution \(P(W)\) approaches its Gaussian reference \(P_G(W)\) as monitoring becomes stronger, consistent with the behavior shown in Fig.~\ref{Fig1}. Parameters: \(L=60\), averaged over \(500\) stochastic realizations.}
\label{Fig2}
\end{figure}
As shown in Fig.~\ref{Fig2}, \(D_{\mathrm{tail}}^{(2\sigma)}\) decreases with increasing cumulative monitoring strength \(\gamma t\). This behavior is consistent with the evolution of \(P(W)\) shown in Fig.~\ref{Fig1}: weak monitoring leaves visible non-Gaussian peak structures, whereas stronger monitoring progressively suppresses these features through measurement-induced dephasing. As a result, the Gaussian reference becomes increasingly accurate, including in the tail region. The decrease of \(D_{\mathrm{tail}}^{(2\sigma)}\) therefore provides a quantitative measure of the crossover of \(P(W)\) toward an effectively Gaussian form.\\~\\
Next, we compute the moments of $P(W)$. To do so, we start from differentiating the trajectory-resolved characteristic function $\mathcal{G}^{(r)}(u)$ in Eq.~(\ref{eq:single_traj_G_fullcorr}) at $u=0$ to obtain
\[\langle W\rangle_r=i\,\partial_u \ln \mathcal{G}^{(r)}(u)\big|_{u=0},
\qquad
\sigma_{W,r}^2=-\partial_u^2 \ln \mathcal{G}^{(r)}(u)\big|_{u=0}.\]
Using the full-correlation generating function, the trajectory-resolved
mean work is
\begin{equation}
\langle W\rangle_r=-E_i^0
+
\sum_{\nu=1}^{2L}
\mathcal{E}_\nu\,
\widetilde{\mathcal C}^{(r)}_{\nu\nu},
\label{eq:mean_work_traj}
\end{equation}
where \(\mathcal{E}_\nu\) are the final BdG energies and
\(\widetilde{\mathcal C}^{(r)}\) is the generalized correlation matrix
in the final BdG basis. The corresponding trajectory-resolved variance is
\begin{equation}
\sigma_{W,r}^2
=
\sum_{\nu=1}^{2L}
\mathcal{E}_\nu^2\,
\widetilde{\mathcal C}^{(r)}_{\nu\nu}
-
\sum_{\nu,\mu=1}^{2L}
\mathcal{E}_\nu\mathcal{E}_\mu
\left|
\widetilde{\mathcal C}^{(r)}_{\nu\mu}
\right|^2 .
\label{eq:var_work_traj}
\end{equation}
After averaging over \(R\) stochastic trajectories, the mean work is
\begin{equation}
\langle W\rangle=\frac{1}{R}\sum_{r=1}^{R}\langle W \rangle_r,
\label{eq:mean_work_ensemble}
\end{equation}
and the total work variance is
\begin{equation}
\sigma_W^2=\frac{1}{R}\sum_{r=1}^{R}{\sigma_{W,r}^2}
+\frac{1}{R}\sum_{r=1}^{R}
\left(
\langle W\rangle_r-\langle W\rangle
\right)^2
.
\label{eq:var_work_ensemble}
\end{equation}
Thus the total variance contains both the quantum variance within each Gaussian trajectory and the trajectory-to-trajectory fluctuations of the mean work.\\~\\
We plot both quantities separately as functions of $\gamma t$ and $h$. Fig.~\ref{Fig3}(a) shows the dependence of $\langle W \rangle$ on the monitoring strength \(\gamma t\) at fixed \(h=0.15\). At \(\gamma t=0\), the system evolves trivially with respect to the final Hamiltonian, and the work distribution is sharply localized around zero. Consequently both \(\langle W\rangle\) and \(\sigma_W^2\) vanish. Once monitoring is switched on, the local quantum jumps act as a source of measurement-induced excitations in the final Bogoliubov basis. This produces the rapid increase of \(\langle W\rangle\) at small \(\gamma t\). At larger monitoring strength, the mean work approaches a plateau. This saturation indicates that the monitored dynamics has reached a regime in which the final-basis occupations generated by the jump process no longer change appreciably with additional monitoring time. In Fig.~\ref{Fig3}(b), we plot the variance $\sigma^2_{W}$, given in Eq.~(\ref{eq:var_work_ensemble}), as a function of $\gamma t$. The first term is the intrinsic quantum variance within an individual Gaussian trajectory, whereas the second term measures the spread of the trajectory-resolved mean work over different jump records. At small but nonzero \(\gamma t\), the ensemble is highly heterogeneous: different trajectories contain different jump histories. This makes the trajectory-resolved mean work \(\langle W\rangle_r\) broadly distributed and produces the initial enhancement of \(\sigma_W^2\) for weak-to-intermediate monitoring. At larger \(\gamma t\), most trajectories contain many jumps and are governed by the typical monitored dynamics. The mean work has then already saturated, and the differences between trajectory-resolved mean works become less pronounced. Consequently, the variance decreases from its maximum and approaches a finite plateau.
\\~\\
Figs.~\ref{Fig3}(c) and ~\ref{Fig3}(d) show the $h$-dependence of the work cumulants at fixed \(\gamma t=4\). For small and intermediate \(h\), the mean work varies only weakly, and gradually decreases at larger \(h\). For small \(h\), the system remains deep in the ordered regime, so changing \(h\) only weakly modifies the final BdG spectrum and the occupations generated by the jump dynamics. As a result, the average final energy, and hence \(\langle W\rangle\), changes slowly. At larger \(h\), the transverse field has a stronger effect on the quasiparticle spectrum and on the final Bogoliubov basis. The same monitoring protocol then redistributes the final-basis occupations so that their net energy-weighted contribution is smaller, leading to the observed reduction of \(\langle W\rangle\). The variance, on the other hand, increases with $h$, as shown in Fig.~\ref{Fig3}(d). This indicates that increasing the transverse field makes the final energies produced by different jump records more broadly distributed. This broadening can occur even when the average work is nearly constant or decreases, because \(\langle W\rangle\) is only a linear average, whereas \(\sigma_W^2\) is sensitive to the separation between different energy outcomes and to the off-diagonal correlations in the final Bogoliubov basis.

\onecolumngrid
\begin{center}
\includegraphics[width=0.85\textwidth]{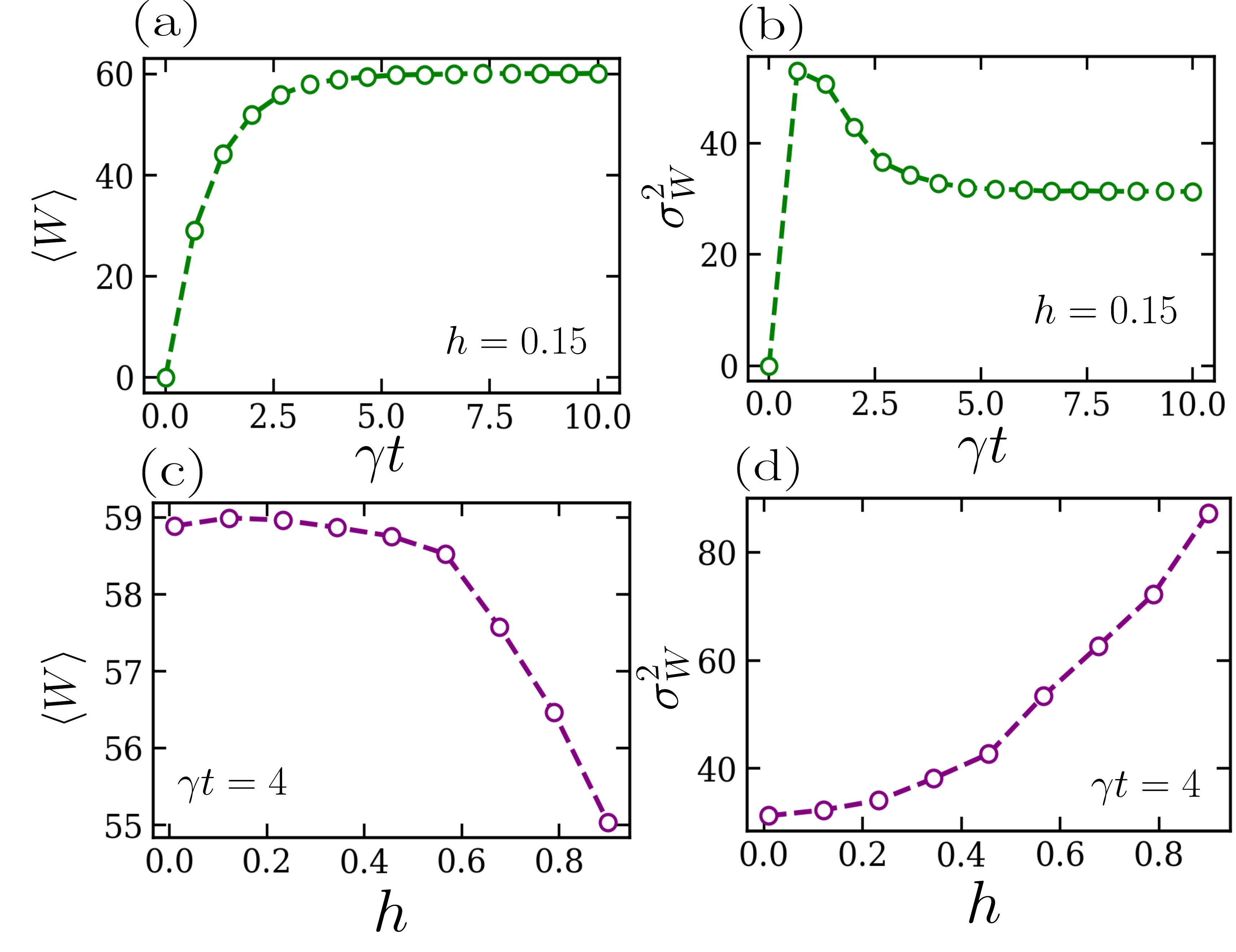}
\captionof{figure}{Work cumulants under stochastic jump dynamics with no quench, $H_i = H_f$, for a system size $L=60$, averaged over $500$ stochastic realizations. (a) Mean work \(\langle W\rangle\) as a function of the cumulative monitoring strength \(\gamma t\) at fixed transverse field \(h=0.15\). (b) Corresponding work variance \(\sigma_W^2\). The mean work increases rapidly and then saturates, whereas the variance shows a pronounced maximum before relaxing to a plateau. (c) Mean work as a function of the transverse field \(h\) at fixed \(\gamma t=4\). The mean work varies only weakly at small and intermediate \(h\), and decreases at larger \(h\). (d) Corresponding work variance, showing that increasing \(h\) broadens the distribution of final-energy outcomes even when the average work is nearly constant or decreasing.}
\label{Fig3}
\end{center}
\twocolumngrid

\section{Controlled Jump Protocol and Light-Cone Effects}
\label{deterministic}
To isolate the impact of inter-jump correlations on work statistics, we introduce a controlled measurement protocol in which local projections onto a definite transverse spin state are deterministically enforced at specified times and sites, rather than emerging stochastically from continuous monitoring. This setup allows us to precisely tune the spatial and temporal separation between jumps, enabling a direct probe of how causality, governed by the light-cone effect associated with the propagation of quasiparticles, impacts the additivity of energetic contributions from multiple jumps. We consider the same transverse-field Ising chain in Eq.~(\ref{QIC}) starting from its ground state with energy $E^{0}_{i}$. Since no quench is performed ($H_i = H_f = H_{\rm QI}$), the work $W$ coincides with the excess energy injected into the system relative to $E^{0}_{i}$, computed via the two-point measurement scheme as the difference between final and initial projective energy measurements. The protocol is defined by fixing in advance a sequence of jump events, specified by the pairs $\{(i_j, t_j)\}_{j=1}^{M}$, where $t_j$ is the time and $i_{j}$ is the site of the $j$-th jump. Between successive jumps, the state evolves unitarily under $H_{\rm QI}$. At each selected time $t_j$ and site $i_{j}$, we impose the outcome \(+1\) of a projective measurement of $\sigma^z_{i_j}$. The corresponding projection operator is \(P^{(+)}_{i_j}\equiv(\mathbb{1}+\sigma^z_{i_j})/2\). Accordingly, denoting the states immediately before and after the \(j\)-th jump by \(|\psi(t^{-}_j)\rangle\) and \(|\psi(t^{+}_j)\rangle\), respectively, the state is updated as
\[|\psi(t^{+}_j)\rangle=\frac{P^{(+)}_{i_j}|\psi(t^{-}_j)\rangle}{\sqrt{\langle \psi(t^{-}_j)| P^{(+)}_{i_j}|\psi(t^{-}_j)\rangle}}.\]
Each such projection injects energy by locally disturbing the state and thereby creating quasiparticle excitations. These excitations propagate ballistically, with maximum quasiparticle group velocity $v_{\rm max} \approx 2J \min(1, h)$ in the thermodynamic limit~\cite{Calabrese_2012}.
The Gaussian form of the state is maintained throughout, allowing efficient numerical simulations through updates of the two-point fermion correlation matrix.\\~\\
\begin{figure}[h]
\centering
\includegraphics[width=0.75\linewidth]{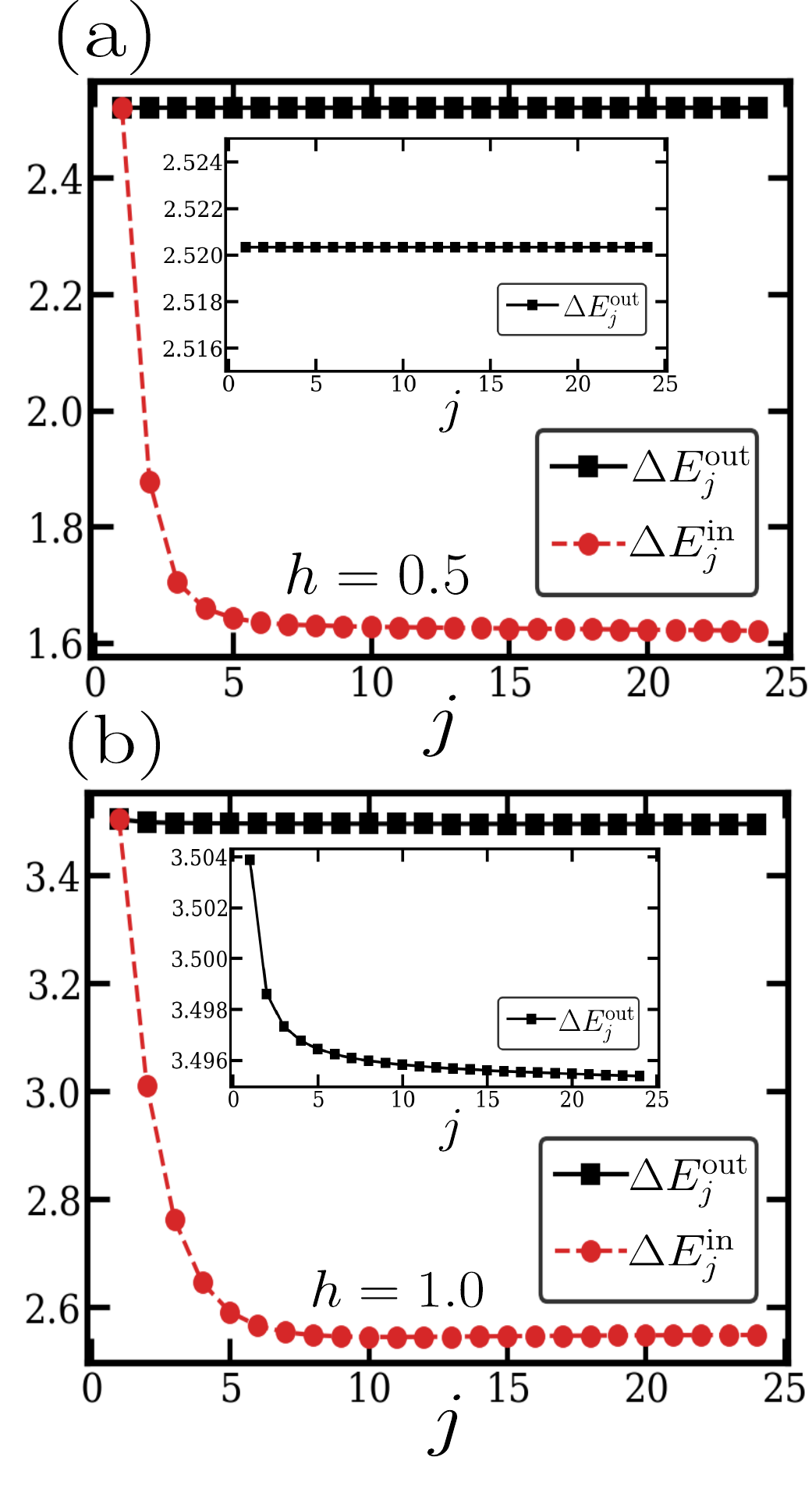}
\caption{Per-jump energy increment $\Delta E_{j}$ as a function of jump index $j$ for system size $L=1024$. (a) In the gapped ferromagnetic phase at $h=0.5$, $\Delta E^{\rm out}_{j}$ remains constant with respect to $j$, while $\Delta E^{\rm in}_{j}$  drops during the first few jumps and then saturates. Inset: zoomed view of $\Delta E^{\rm out}_{j}$ confirming its $j$-independence. (b) At the critical point $h=1.0$, algebraic correlations render even the outside-cone $\Delta E^{\rm out}_{j}$ weakly $j$-dependent, as revealed in the inset, whereas the inside-cone $\Delta E^{\rm in}_{j}$ still decays and saturates but more slowly.}
\label{Fig4}
\end{figure}
We distinguish two regimes based on whether consecutive jumps are causally connected or disconnected. The light cone emerging from a jump at site $i$ and time $t$ includes all sites $j$ where $|i-j| \leq v_{\rm max} |t'-t|$, within which disturbances can propagate and interact~\cite{bravyi}.  Here, $ t' $ denotes a subsequent time at which we evaluate the propagation of the perturbation caused by a jump, relative to the initial jump time $t$. Jumps outside this light cone (spacelike separated) remain approximately causally disconnected, leading to the exponential decay of correlations beyond the cone, as established by the Lieb-Robinson bound. In contrast, jumps inside the light cone (timelike separated) allow for interference through overlapping quasiparticle trajectories, resulting in correlated and non-additive energy transfers.\\~\\
We quantify these effects through the per-jump energy increment \[\Delta E_j=\left\langle \psi(t_j^{+}) \right|
\hat{H}_{\mathrm{QI}}
\left| \psi(t_j^{+}) \right\rangle
-
\left\langle \psi(t_j^{-}) \right|
\hat{H}_{\mathrm{QI}}
\left| \psi(t_j^{-}) \right\rangle,\] 
defined as the energy change associated with the $j$-th jump. For jumps outside the light cone, $\Delta E_j^{\text{out}}$ remains essentially constant and independent of $j$, reflecting the absence of causal overlap: each jump injects a fixed amount of energy by disrupting local correlations and generating quasiparticles that propagate without interfering with those generated by other jumps. This leads to a linear growth in the average work $\langle W \rangle$ with the number of jumps $j$. For jumps inside the light cone, $\Delta E_j^{\text{in}}$ drops sharply with $j$ for early steps and then saturates, exhibiting non-additive behavior due to causal interactions: earlier jumps modify the local quasiparticle content, dephasing the state and partially projecting it toward the measurement eigenbasis (Zeno-like effect), which reduces the energy that can be injected by subsequent jumps. This produces a transiently sublinear buildup of $\langle W \rangle$, followed by linear growth with a reduced slope once $\Delta E_j^{\text{in}}$ reaches its plateau. In Fig.~\ref{Fig4}(a), we show the behavior of $\Delta E_j$ for the non-critical regime. For $h = 0.5$, the system exhibits short-range correlations. Here, $\Delta E_j^{\text{out}}$ is constant across $j$, confirming negligible dynamical overlap outside the light cone (see the inset). Meanwhile, $\Delta E_j^{\text{in}}$ shows the characteristic “drop-then-plateau” behavior, caused by overlapping quasiparticles and local dephasing from repeated projections. At criticality, the chain is gapless, and equal-time correlations have algebraic tails. This long-range baseline makes even spacelike-separated regions weakly sensitive to prior jumps. Accordingly, in Fig.~\ref{Fig4}(b), although $\Delta E_j^{\text{out}}$ appears flat at first glance, a zoomed-in view (as shown in the inset) reveals non-constant values with subtle variations across $j$. Inside the cone, $\Delta E_j^{\text{in}}$ still decays and saturates, but the decay is slower than in the gapped case, reflecting the critical correlations. Fig.~\ref{Fig4} naturally explains how the average work builds up in the controlled-jump protocol. Since the total injected energy is the sum of contributions from all imposed jumps, the average work after \(M\) jumps is $\langle W \rangle = \sum_{j=1}^{M}\Delta E_j$. The behavior of $\langle W \rangle$ is  therefore determined by how $\Delta E_{j}$ changes with $j$. A constant $\Delta E_{j}$ outside the light cone therefore implies a linear growth of $\langle W \rangle$ with increasing $M$, while inside the light cone, where $\Delta E_{j}$ decreases with $j$ before saturating, $\langle W \rangle$ initially grows sublinearly and subsequently crosses over to linear growth. This should be contrasted with the stochastic weak measurement protocol discussed earlier. There, with increasing $\gamma t$, continuous weak monitoring repeatedly drives the state toward the measurement basis and the dynamics becomes progressively less effective at creating new excitations. As a result, the extra work added at later times becomes small, so $\langle W \rangle$ eventually saturates, which is naturally interpreted as a Zeno-like effect.
\onecolumngrid
\begin{center}
\includegraphics[width=1.0\linewidth]{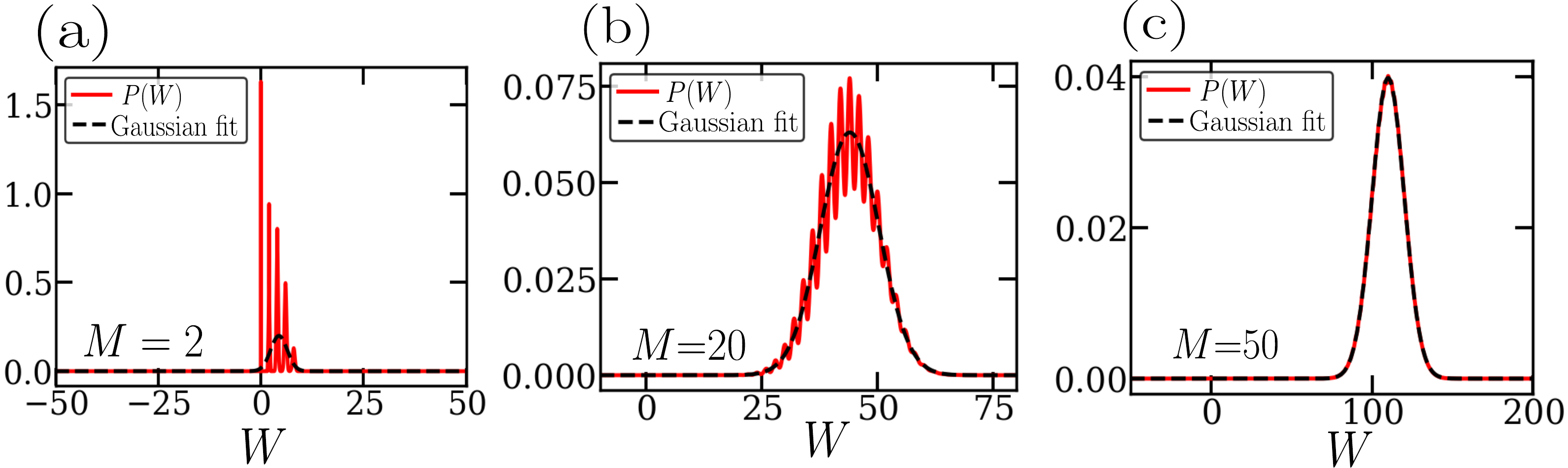}
\captionof{figure}{Work distribution $P(W)$ in the controlled jump protocol with equally spaced jumps in space and time, illustrating the progression toward Gaussianity with an increasing number of jumps $M$ for system size $L=300$, $h=0.2$, and total time $t=1$. (a) For $M=2$, $P(W)$ displays a dominant central peak at $W=0$ indicative of minimal energy injection, accompanied by multiple discrete peaks at $W>0$, reflecting a small set of coherent excitations and preserved coherences. (b) For $M=20$, the distribution develops a bell-shaped core but retains residual spikes from incomplete dephasing. (c) For $M=50$, $P(W)$ approaches a fully smooth profile that closely overlaps with the Gaussian fit with the same mean and variance over the plotted range.}
\label{Fig5}
\end{center}
\twocolumngrid
Using the same protocol, we also examined how the features of the full work distribution $P(W)$ vary with the number of jumps. Also in this case, as the number of jumps $M$ increases in the controlled jump protocol with equally spaced intervals in space and time, the work distribution $P(W)$ undergoes a progressive transformation toward Gaussianity, driven by the cumulative effects of jump-induced dephasing and energy increments. As illustrated in Fig.~\ref{Fig5}(a) for $M=2$, $P(W)$ exhibits a dominant central peak at $W=0$, indicative of minimal energy injection with limited disruptions, together with multiple discrete peaks at $W>0$ that arise from a small set of quasiparticle excitations and residual coherences preserved between the jumps. When the number of jumps is increased to $M=20$, as shown in Fig.~\ref{Fig5}(b), $P(W)$
already develops a bell shape in its core, though it still features residual spikes arising from persistent inter-jump correlations and partial dephasing. At the larger value $M=50$, as depicted in Fig.~\ref{Fig5}(c), $P(W)$ approaches a smooth profile well fitted by a Gaussian, as frequent projections suppress higher-order cumulants and approximate CLT behavior for the sum of weakly correlated yet mixing increments. In practice, this manifests as a crossover from a comb-like non-Gaussian $P(W)$ at $M = 2$ to a near-Gaussian form at large $M$, consistent with the two-point measurement framework and CLTs for additive functionals in quantum processes~\cite{kipnis}.\\~\\
While light-cone effects introduce correlations that manifest in the per-jump energy increments $\Delta E_j$, leading to non-additive behavior inside the cone and independence outside, these correlations primarily influence the mean, $\langle W \rangle = \sum_j \langle \Delta E_j \rangle$, and the variance $\sigma_W^2$ of the work distribution $P(W)$, without altering its overall Gaussian shape. To demonstrate this, Fig.~\ref{Fig6} compares $P(W)$ for larger number of jumps $M=50$ for two deterministic schedules that differ only in whether successive jumps lie inside or outside each other’s light cone.
\begin{figure}[h!]
\centering
\includegraphics[width=0.8\linewidth]{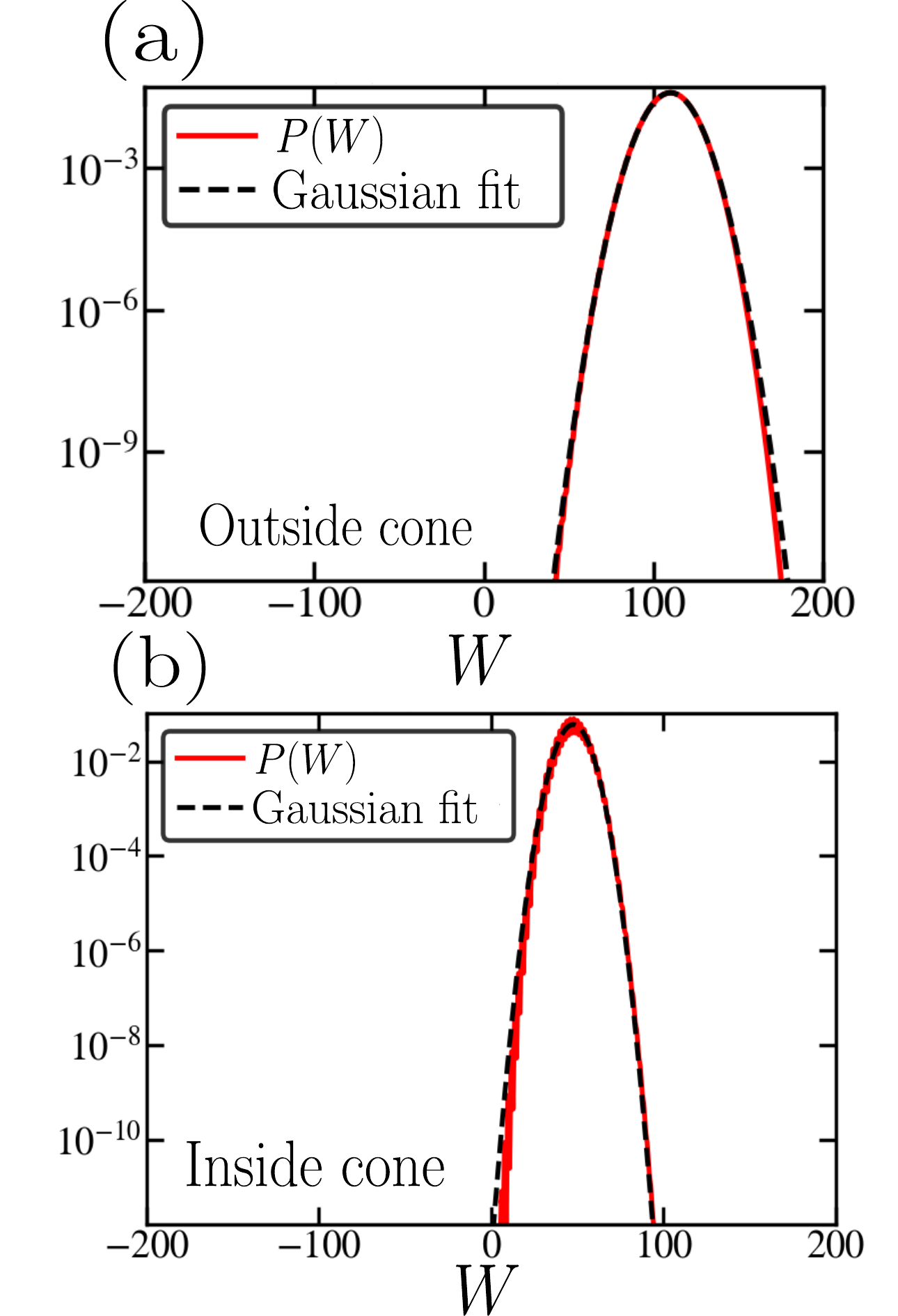}
\caption{
Work distribution \(P(W)\) for \(M=50\) deterministic jumps in the controlled protocol, plotted on a logarithmic scale and compared with Gaussian references constructed from the same mean and variance. (a) Outside-cone sequence: successive jumps are approximately causally disconnected, producing a distribution centered at larger \(W\) with a broader width. (b) Inside-cone sequence: correlated non-additive increments shift the distribution to smaller \(W\) and reduce its width. In both cases the core is close to Gaussian at large \(M\), while the inside-cone case retains weak finite-\(M\) tail structures. Parameters: \(L=300\), \(h=0.2\), total time \(t=1\).}
\label{Fig6}
\end{figure}
Outside the light cone [Fig.~\ref{Fig6}(a)], successive jumps are approximately causally disconnected. Their energetic contributions therefore add nearly independently, producing a work distribution centered at larger \(W\) and with a broader width. Inside the light cone [Fig.~\ref{Fig6}(b)], later jumps act on regions already modified by earlier projections. The corresponding energy increments are therefore correlated and non-additive, leading to a smaller mean work and a narrower distribution. Despite this difference in the first two cumulants, both $P(W)$ profiles remain asymptotically Gaussian at larger $M$, as indicated by closely matching Gaussian fits (black dashed lines) and negligible tail deviation in the logarithmic scale. However, the inside-cone distribution can retain weak residual structures on the lower-\(W\) side. We interpret these features as finite-\(M\) remnants of correlated jump dynamics.
\\~\\
This behavior is consistent with the two-point measurement framework, where $P(W)$ is the inverse Fourier transform of the single-trajectory characteristic function $\mathcal{G}(u)$; for many jumps, a central-limit mechanism suppresses higher cumulants so that the shape remains Gaussian while its center and width reflect the underlying increment statistics. In other words, $P(W)$ aggregates $\langle W \rangle = \sum_{j=1}^M \Delta E_j$ into a marginal distribution dominated by the first two cumulants at large $M$, making it insensitive to detailed spatiotemporal dependence. This implies that varying correlation structures among energy increments can produce observationally equivalent work distributions~\cite{nelsen2006copulas}.\\~\\
\section{WORK STATISTICS UNDER QUENCH AND STOCHASTIC JUMP DYNAMICS}
\label{quench_dynamics}
In this section, we extend our analysis to include a sudden unitary quench with continuous measurements via stochastic quantum jumps. We explore how coherent driving competes and combines with measurement-induced dephasing and mixing, altering the shape of the work distribution $P(W)$.\\~\\
We first recall the case of an isolated quench, which involves an abrupt change in the Hamiltonian parameter, here, the transverse field $h$, from an initial value $h_i$ to a final value $h_f$ at time $t=0$, followed by evolution under the post-quench Hamiltonian $H_f$. This protocol introduces coherent quasiparticle excitations through the mismatch between the pre- and post-quench eigenbases, resulting in sharp, model-dependent patterns in the work statistics, even without  monitoring~\cite{silva}.
\onecolumngrid
\begin{center}
\vspace{3mm}
\includegraphics[width=1.0\textwidth]{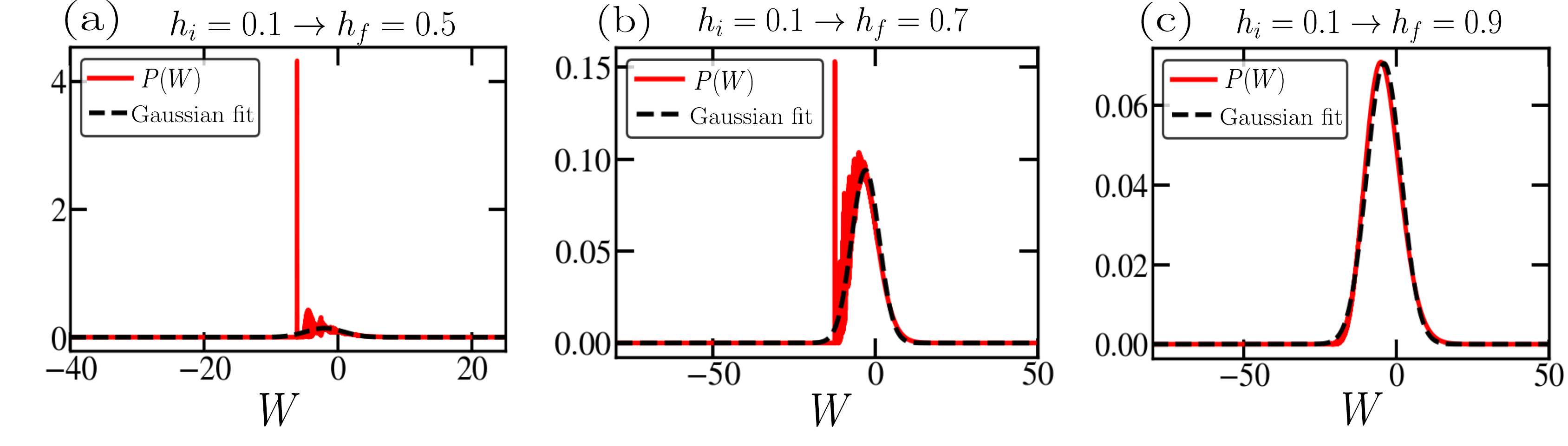}
\captionof{figure}{\label{Fig7}Work distribution $P(W)$ for an isolated sudden quench of the transverse-field
Ising chain from $h_i = 0.1$ to (a) $h_f = 0.5$, (b) $h_f = 0.7$, and (c) $h_f = 0.9$. (a) For the small quench $h_f = 0.5$, $P(W)$ is extremely sharp and strongly non-Gaussian, with a dominant delta-like peak at the minimum work $W_{\rm min}=E_0^f-E_0^i <0$ and only a very weak side structure, reflecting that the initial state is still close to an eigenstate of $H_f$. (b) For the intermediate quench $h_f = 0.7$, $P(W)$ splits into a narrow delta-like spike at \(W_{\rm min}\) of weight $|\langle 0_f | 0_i \rangle|^2$ from trajectories ending in the post-quench ground state, plus a broadened component from Bogoliubov quasiparticle excitations over many modes, whose envelope is approximately Gaussian and whose residual spikes are finite-size precursors of continuum edge singularities. (c) For the strong quench $h_f = 0.9$, the ground-state overlap is negligible and the delta-like contribution disappears; many Bogoliubov pairs are excited over a wide momentum range, and their sum yields a smooth, bell-shaped $P(W)$.}
\label{Fig7}
\end{center}
\twocolumngrid
Fig.~\ref{Fig7} illustrates how the work distribution $P(W)$ changes as we increase the value of the final field $h_f$ in the transverse-field Ising chain, starting from the same initial field $h_i=0.1$ and jumping to $h_f=0.5,\,0.7,\,\text{and}\,0.9$. For the smallest final-field value, $h_f=0.5$ [Fig.~\ref{Fig7}(a)], the distribution is extremely sharp and strongly non-Gaussian, dominated by a delta-like contribution at the minimum work $W_{\rm min}=E_0^f-E_0^i$, which is negative for this quench. This is the regime where the post-quench Hamiltonian is only slightly different from the pre-quench one, so the initial state is still very close to an eigenstate of the new Hamiltonian. When we increase the final field to $h_f=0.7$, more energy is injected, and the work distribution $P(W)$ develops two clearly identifiable contributions, as depicted in Fig.~\ref{Fig7}(b). First, a narrow delta spike with weight $|\langle 0_f|0_i \rangle|^2$ reflects the probability that the post-quench energy measurement finds the system already in the final ground state. Second, a broadened component comes from exciting quasiparticles across many momentum modes. 
Because the total work is the sum of many  contributions, this broadened part has an approximately Gaussian envelope, while the visible distinct spikes are finite-size precursors of threshold onsets in the continuum and the associated edge singularities, which are hallmarks of integrable quenches in the transverse-field Ising model~\cite{silva,smacchia2013}.
For the strong quench $h_f=0.9$, the ground state of the final Hamiltonian differs substantially from that of the initial Hamiltonian. As a result, their overlap is very small, so the probability of projecting onto the post-quench ground state becomes negligible and the delta-like contribution is no longer visible. At the same time, the quench creates excitations over a broad range of momentum modes, so the work distribution is dominated by the sum of many mode contributions. This makes the central part of $P(W)$ very close to Gaussian, and any residual fine structure is essentially washed out, leaving a smooth bell-shaped curve (see Fig.~\ref{Fig7}(c)).\\~\\
\begin{figure}[h!]
\centering
\includegraphics[width=0.88\linewidth]{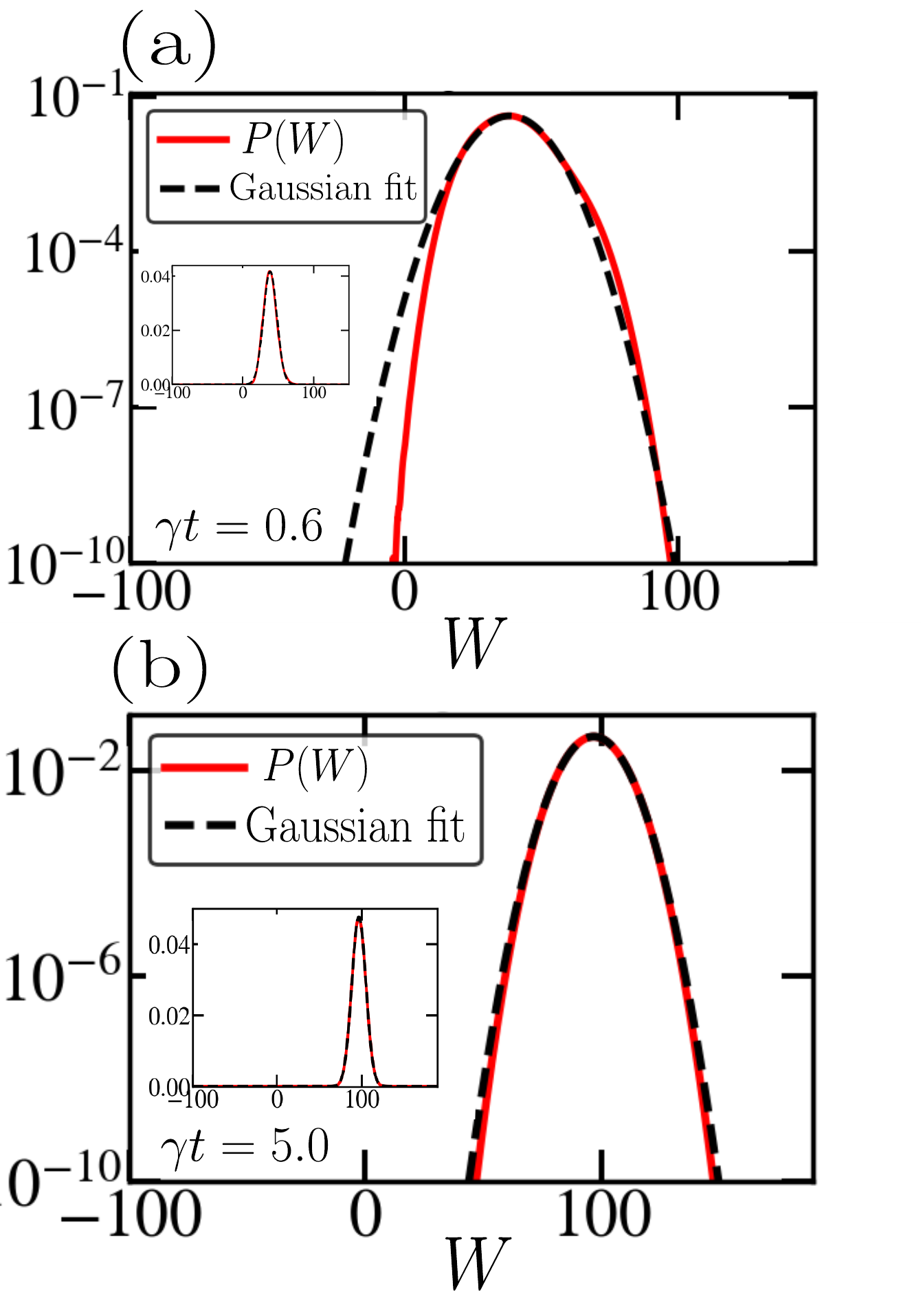}
\caption{Work distribution $P(W)$ under stochastic quantum-jump dynamics combined with a sudden quench in the monitored transverse-field Ising chain from $h_i = 0.1$ to $h_f = 0.5$, for system size $L = 100$, averaged over 200 realizations. The main panels show $P(W)$ on a logarithmic scale to highlight tail behavior; insets display the corresponding cores on a linear scale. (a) Weak monitoring ($\gamma t=0.6$): rare jumps suppress the spikes from isolated-quench distribution in Fig.~\ref{Fig7}(a), yielding an apparently Gaussian core on a linear scale due to central-limit aggregation, but sub-Gaussian tails on a logarithmic scale decay faster than the Gaussian fit, reflecting bounded fluctuations and residual correlations from incomplete dephasing. (b) Strong monitoring ($\gamma t = 5.0$): frequent jumps strongly enhance dephasing and suppress higher cumulants, yielding a fully Gaussian $P(W)$ that matches the fit across both the bulk and the tails.}
\label{Fig8}
\end{figure}
Let us now investigate how incorporating generalized measurements, which induce stochastic quantum jumps after the quench, modifies the resulting work distribution. Fig.~\ref{Fig8} illustrates how continuous monitoring progressively makes the work statistics more Gaussian associated with the quench from $h_i = 0.1$ to $h_f = 0.5$, of Fig.~\ref{Fig7}(a). In the absence of measurements, that quench produces a highly structured distribution: a sharp contribution near the post-quench ground state plus a broadened part built from coherent Bogoliubov excitations over many modes, whose envelope is only approximately Gaussian and whose fine structure encodes the integrable quench physics. As in the previous cases, when weak monitoring ($\gamma t = 0.6$) is turned on, rare jumps partially disrupt coherent quasiparticles, suppressing the isolated-quench spikes and yielding an apparently Gaussian core on a linear scale (see inset of Fig.~\ref{Fig8}(a)) due to the central-limit aggregation of mode-resolved energy increments;  however, the logarithmic-scale plot in Fig.~\ref{Fig8}(a) reveals sub-Gaussian tails that decay faster than a true Gaussian fit, indicative of bounded fluctuations and residual correlations from incomplete dephasing, akin to the intermediate regime in unquenched dynamics (see Fig.~\ref{Fig1}(c)) where higher cumulants persist. By contrast, under strong monitoring ($\gamma t = 5.0$), as shown in Fig.~\ref{Fig8}(b), jumps become more frequent, higher cumulants are strongly suppressed, and the quench-induced structures are washed out. The work is then dominated by the additive statistics of many weakly correlated increments, and $P(W)$ approaches an almost fully Gaussian form even on a logarithmic scale, consistent with a central-limit mechanism in a monitored many-body system~\cite{plenio,hoeffding,kipnis,jones,kurtz}.

\section{Conclusions and Outlook}
\label{conclusion}
In this work, we studied work statistics in a monitored transverse-field Ising chain under stochastic quantum jumps, controlled jump sequences, and sudden quenches using a fermionic Gaussian-state framework. Within the two-point measurement scheme, we derived a trajectory-resolved generating function for work in the presence of generalized measurements and used it to compute work distributions in an extended monitored many-body system beyond single- or few-level platforms.\\~\\
For stochastic jump dynamics without a quench, we found that the work distribution crosses over from a discrete comb-like structure at weak monitoring to an essentially Gaussian form at strong monitoring, with sub-Gaussian tails in the intermediate regime. We also showed that the mean work and its variance display a growth-and-saturation crossover with increasing monitoring strength. Moreover, to isolate the role of inter-jump correlations, we analyzed deterministic jump sequences applied at controlled positions and times. Outside the Lieb-Robinson light cone, the energy increment per jump remains essentially constant, leading to linear growth of the average work with the number of jumps. Inside the light cone, the per-jump increment decreases and then saturates, producing a transiently sublinear crossover followed by linear growth with a reduced asymptotic slope. At criticality, even spacelike-separated jumps acquire a weak dependence on the jump index, reflecting long-range correlations. We further found that, although these correlations strongly affect the buildup of the mean work, the full work distribution becomes nearly Gaussian once many jumps are accumulated. Finally, for monitored quenches, we showed that continuous observation progressively suppresses the fine structure of the isolated-quench work distribution and drives it toward Gaussian behavior. At sufficiently strong monitoring, the statistics are dominated by the accumulation of many weakly correlated measurement-induced energy increments superimposed on the quench-induced excitations.\\~\\
These results show that work statistics offers insights into monitored many-body dynamics beyond the average energy change alone. By resolving the distribution and the moments of work across individual trajectories, this approach clarifies how measurement backaction, coherent evolution, and spatiotemporal correlations combine to shape the nonequilibrium dynamics of the system. More broadly, our work demonstrates that work statistics provides a useful perspective on monitored many-body dynamics in a setting where most previous studies have focused primarily on information-theoretic observables. We hope that this work will motivate further studies of thermodynamic observables in monitored many-body systems and help strengthen the connection between quantum thermodynamics and measurement-driven many-body dynamics.
\section*{Acknowledgments}
A. S. and M. M. acknowledge the support of the PNRR MUR grant PE0000023-NQSTI. A. S. acknowledges the support of the project ``Superconducting quantum-classical linked computing systems (SuperLink)'' within the framework of QuantERA2 ERANET COFUND in Quantum Technologies.\\~\\
\section*{DATA AVAILABILITY}
The numerical data and codes supporting the findings of this
study are available from the corresponding authors upon
reasonable request.

\label{quench}
\label{conclusion}

\bibliography{WSjump_refs}

\end{document}